**Replicability and the public/private divide**

Dear Sir,

In a recent letter, Carlos Vilchez-Román criticizes Bornmann *et al*. (2015) for using data which cannot be reproduced without access to an in-house version of the Web-of-Science (WoS) at the Max Planck Digital Libraries (MPDL, Munich). We agree with the norm of replicability and therefore returned to our data. Is the problem only a practical one of automation or does the in-house processing add analytical value to the data?

*Replicability of our study*

Vilchez-Román (*personal communication*; August 15, 2015) tried to replicate the number of top-1% most-highly cited publications of Brazil during the period 1990-2010. This number is provided in Table 1 of our paper (at p. 1509) as 1,309 papers corresponding to 0.5% of the world total based on integer counting and normalization for document types (articles, reviews, and letters), publication years, and WoS Subject Categories (WCs). Integer counting is also used when searching online.

For example, searching for "PY=2012" and only articles (as a specific document type), we retrieved 1,320,618 records on September 10, 2015; of these articles 36,927 had at least one institutional address in Brazil (2.80%). Using the method specified by Ahlgren *et al*. (2014)—that is, sorting the retrieval from most to least cited within WoS—one can find that the top-1% most-highly cited of the reference set ("the world") are cited 53 or more times. In the Brazilian sample, 195 records have 53 or more citations; this is 0.53% of the reference set.

One can repeat this for every publication year; but one would have to repeat the analysis also for all 250 WCs. For "Plant Sciences," for example, the Brazilian contribution is 1.58%. However, the results of individual runs using different WCs can no longer be aggregated to a result for "Brazil" because most journals—and therefore the publications within them—are assigned to more than a single WC (Leydesdorff & Rafols, 2009, p. 350). Using integer counting, the aggregation to a result for "Brazil" would lead to double counting both in the Brazilian and the reference set. One thus needs a scheme (fractional counting) for weighing multiple WCs attributed to a single paper.

*A professional divide?*

MPDL uses a scheme of fractional counting different from, for example, InCites of Thomson Reuters or CWTS in Leiden. Routinization of these (and similar) procedures classifies a number of institutions as professional centers producing algorithmic results which cannot be reproduced without access to their in-house database of preprocessed data (cf. Hicks *et al*., 2015). The algorithms, thesauri, etc., function as the intellectual capital of these centers as quasi-firms, and are therefore not freely accessible.

Is the newly emerging situation in any sense different from a further professionalization of the field? Access to WoS also requires subscription. In our opinion, a political economy of science indicators has in the meantime emerged with a competitive dynamic that affects the intellectual organization of the field. On the research side, inequalities in access are reinforced by this public/private divide. The publications of the professional centers legitimize their competitive advantage on the market of policy reports and management tools.

The competitive structure tends to become oligopolistic: research is increasingly concentrated in leading centers because of the costs involved in meeting the professional standards. Note that the leading centers can be small-sized: SciTech Strategies, for example, consists of only three researchers. The competitive edge is not size, but investments in advanced knowledge. This socio-economic context of a knowledge-based economy may make it necessary to redefine "replicability" as well as other norms of science in relation to changing patterns of specialization (Ziman, 2000).

Loet Leydesdorff *,[a], Caroline Wagner [b], and Lutz Bornmann [c]

* Corresponding author;

[a] Amsterdam School of Communication Research (ASCoR), University of Amsterdam, PO Box 15793; 1001 NG Amsterdam, The Netherlands; email: loet@leydesdorff.net

[b] Battelle Center for Science & Technology Policy, John Glenn School of Public Affairs, Ohio State University, Columbus, OH (USA); email: wagner.911@osu.edu

[c] Division for Science and Innovation Studies, Administrative Headquarters of the Max Planck Society, Hofgartenstr. 8, 80539 Munich, Germany; email: bornmann@gv.mpg.de